\documentclass[twocolumn]{aastex62}

\usepackage{natbib}
\usepackage{graphicx}
\usepackage{times}

\begin{document}

\accepted{for publication in ApJ}


\title{PHYSICAL MODELS FOR THE CLUSTERING OF OBSCURED AND UNOBSCURED QUASARS}

\author{Kelly E. Whalen}
\affiliation{Department of Physics \& Astronomy, Dartmouth College, 6127 Wilder Laboratory, Hanover, NH 03755, USA.}

\author{Ryan C. Hickox}
\affiliation{Department of Physics \& Astronomy, Dartmouth College, 6127 Wilder Laboratory, Hanover, NH 03755, USA.}

\author{Michael A. DiPompeo}
\affiliation{Department of Physics \& Astronomy, Dartmouth College, 6127 Wilder Laboratory, Hanover, NH 03755, USA.}

\author{Gordon T. Richards}
\affiliation{Department of Physics, Drexel University, 3141 Chestnut Street, Philadelphia, PA 19104, USA.}

\author{Adam D. Myers}
\affiliation{Department of Physics and Astronomy, University of
Wyoming, Laramie, WY 82071, USA.}

 \shortauthors{Whalen et al.}
 \shorttitle{Physical Models Quasar Clustering}

\begin{abstract}
Clustering measurements of obscured and unobscured quasars show that obscured quasars reside in more massive dark matter halos than their unobscured counterparts.  These results are inconsistent with  simple  unified  (“torus”)  scenarios,  but  might  be  explained  by  models  in  which  the  distribution of obscuring material  depends  on  Eddington  ratio  or  galaxy  stellar  mass.   We  test  these  possibilities  by  constructing simple physical models to compare to observed AGN populations.  We find that previously observed relationships between obscuration  and  Eddington  ratio  or  stellar  mass  are  not  sufficient  reproduce  the  observed  quasar clustering results ($\langle \log M_{\text{halo}}/M_{\odot} \rangle = 12.94 ^{+ 0.10} _{- 0.11}$ and $\langle \log M_{\text{halo}}/M_{\odot} \rangle = 12.49 ^{+ 0.08} _{- 0.08}$ for obscured and unobscured populations, respectively) while maintaining the observed fraction of obscured quasars (30-65$\%$).  This work suggests that evolutionary models, in which obscuration evolves on the typical timescale for black hole growth, are necessary to understand the observed clustering of mid-IR selected quasars.
\\
\end{abstract}


\section{Introduction}
Quasars, the highly luminous subclass of active galactic nuclei (AGN), are among the most energetic objects in the universe, and they are powered by supermassive black holes (SMBHs) that are rapidly accreting matter \citep[e.g.,][]{alex12bhgrowth}. AGN are often characterized in optical observations by the presence of broad emission features in their spectra, as well as a luminous continuum at rest-frame ultraviolet and optical wavelengths \citep[e.g.,][]{bald77luminosityquasars, netz15unification,pado17AGN}. However, there are many observed AGN that appear to lack one or both of these features. Spectropolarimetric measurements have shown astronomers that these ``missing'' features are still present, but these photons have been scattered off of some obscuring material before they were observed \citep[e.g.,][]{anto85blr}. This leads us to understanding that quasars can be classified as either ``obscured'' or ``unobscured.'' Here, we define a quasar as being obscured if it is shielded by a line-of-sight (LOS) column density ($N_{\text{H}}$) of at least $10^{22} \ \text{cm}^{-2}$ \citep[e.g.][] {usma14obscuration, hick18obsc-review}.

The simplest picture of quasar obscuration is that it is an effect due to quasars being randomly oriented relative to an observer. This model of unification by orientation  \citep[e.g.,][]{anto93review,urry95unifiedmodel,netz15unification,ramo17obscuration} suggests that all AGN, including quasars, consist of a SMBH with an accretion disk and an axis-symmetric distribution of dust, also known as a ``dusty torus.'' The non-spherical geometry of the dusty torus can obscure the nucleus of the AGN for some lines-of-sight, meaning that orientation alone could determine whether or not a quasar is obscured to an observer.

Constraints on this unified picture can be obtained through statistical measurements of the properties of large populations of quasars, both obscured and unobscured. A particularly useful measurable property is spatial clustering, which can determine the masses of the dark matter halos that host quasars and their connection to the large-scale environment, independent of the detailed properties of the individual host galaxies which can be difficult to measure for luminous AGN \citep[e.g.][]{conr13quasars, veal14quasars}. Until recently, these measurements have focused on optically-selected unobscured sources or X-ray selected AGN \citep{croo04qso,rich06QLF,myer07clustering,shen09SDSSclustering,ross09SDSScluster,efte15clust}. The dawn of deep, wide mid-infrared (IR) surveys has allowed us to better understand the environments of obscured quasars \citep[e.g.,][]{wern04spitzer,hick07bootes,hick09ir-agn,wrig10wise,krum12clustering,hain14wiseqso,dipo14ir-cluster,dipo16wise-planck,dipo17qsoclust}. With a large sample of mid-IR selected obscured quasars, we can perform statistical analyses to determine if obscured and unobscured quasars are fundamentally different from one another. For unobscured quasars, spatial clustering measurements have shown that their parent dark matter halo masses are roughly constant across a a redshift range of $0 < z < 5$ \citep[e.g.,][]{croo05qso,myer07clustering,shen07highzclust,coil07DEEP2, daan08qsoclust,ross09SDSScluster,hick11bootes_clust,powe18xrayclust}. For obscured quasars selected by the \textit{Wide-field Infrared Survey Explorer} (\textit{WISE}) \citep[e.g.,][]{wrig10wise}, it has been measured that for a given redshift, obscured quasars typically reside in higher mass dark matter halos than their unobscured counterparts \citep[e.g.,][]{hick11bootes_clust, dono14cluster, dipo14ir-cluster,dipo16wise-planck, dipo17qsoclust, powe18xrayclust}. For this paper, we will adopt recent measurements from \citet{dipo17qsoclust} that indicate obscured quasars reside in dark matter halos that have an average mass of $\log M_{\text{halo}}/M_{\odot} = 12.94 ^{+ 0.10} _{- 0.11}$, while unobscured quasars on average reside in dark matter halos of $\log M_{\text{halo}}/M_{\odot} = 12.49 ^{+ 0.08} _{- 0.08}$.\footnote{We note that \citet{dipo17qsoclust} defined quasar obscuration using an optical/mid-IR color cut of $r-W2 = 6$ (Vega) \citep[e.g.,][]{hick07bootes,hick17qsoSED} This cut takes advantage of the fact that obscured and unobscured quasars occupy different parts of SDSS/\textit{WISE} color space \citep[e.g.,][]{hick07bootes}. \citet{hick17qsoSED} showed that SED models are able to predict optical/mid-IR colors for obscured and unobscured quasars that are consistent with observations. The color cut used in \citet{dipo17qsoclust} corresponds to the output of \citet{hick17qsoSED} SED model that assumed $A_{V} = 20$. Based on equation (3) in \citet{drai03dust}, this gives $N_{\text{H}} \sim 3.7 \times 10^{22} \ \text{cm}^{-2}$, which is consistent with our adopted definition of quasar obscuration.} These results provide observational constraints for any model that tries to explain the relationship between obscured and unobscured quasars.

In contrast with the simplest cases of the unified model of AGN,  quasar obscuration may be a phase in an evolutionary scenario that occurs on timescales of roughly a Salpeter (\textit{e}-folding) time for black hole growth at Eddington-limited accretion. This obscuration phase can be associated with dust structures produced during major galaxy mergers \citep[e.g.,][]{silk98galformation,sprin05mergers,hopk06merger,gold12agn-dust, trei12merger, blec18agnmerge}, or it can also be tied to an early phase in a quasar's lifetime at which it is not luminous enough to rid its nucleus of obscuring material \citep[e.g.,][]{hopk08evolve,king10eddington}. The possibility that obscured and unobscured quasars may represent different evolutionary stages allows for them to have different physical properties, such as the masses of their parent dark matter halos. 

Many evolutionary models postulate that as dark matter halos grow, black hole growth lags behind \citep[e.g.,][]{alex08weigh_bh,woo08msigma,korm13coevolution,dipo17model}. As these black holes grow in mass, they transition from an obscured phase to an unobscured phase via radiatively-driven blowout \citep[e.g.,][]{hopk06merger, hopk08evolve}.  \citet{dipo17model} presented a simple evolutionary model in which black hole growth lagged behind galaxy growth. In the \citet{dipo17model} model, the host dark matter halo grows continuously, while the black hole grows in brief episodes. Here, the black hole's change in mass determines the quasar's evolution from obscuration to being unobscured. \citet{hick07bootes,hick11bootes_clust} showed that bolometric luminosities were similar for  populations of obscured and unobscured quasars selected in the mid-IR with \textit{Spitzer Space Telescope} \citep{wern04spitzer}. Since luminosity is just a function of Eddington ratio and black hole mass, assuming similar Eddington ratio distributions implies that both unobscured and obscured quasars of a given luminosity should have the same black hole mass, independent of obscuration. In this model, a black hole will begin to grow if it falls too far off the $M_{*}-M_{\text{BH}}$ relation. As the black hole gains mass, the quasar will become luminous enough to rid its nucleus of some obscuring material, and it will then transition from an obscured to an unobscured stage in its evolution. Because the black hole masses of obscured quasars are similar to that of their unobscured counterparts, their dark matter halo masses are predicted to be larger, which is what is empirically seen.
\begin{figure}
\begin{center}
\includegraphics[width= 0.5\textwidth]{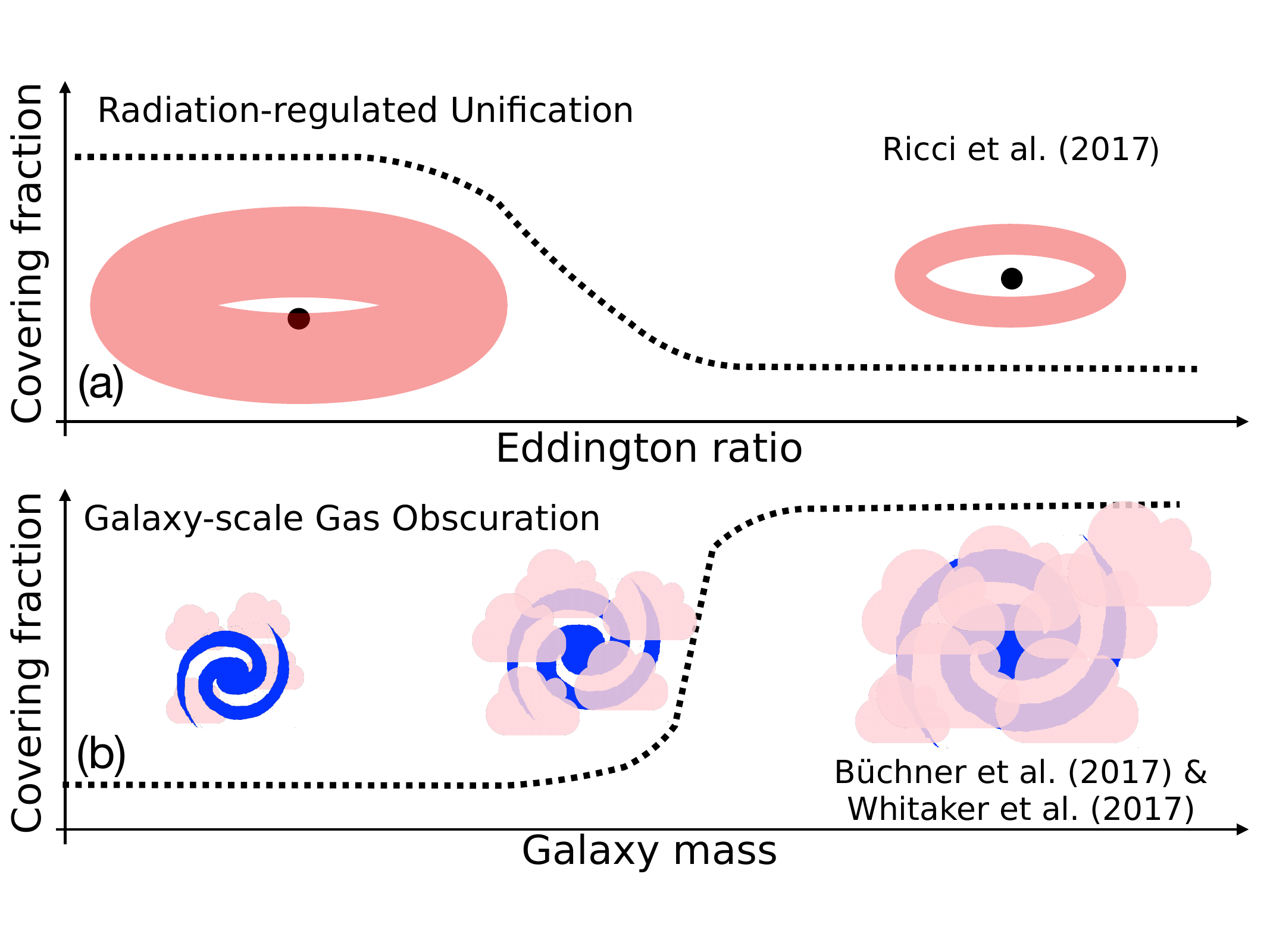}
\end{center}
\caption{\textit{Panel (a)}: Schematic of radiation-regulated unification \citep{ricc17edd}. Studies of X-ray selected AGN show that there is a possible relationship between an AGN's covering fraction and its Eddington ratio. Quasars accreting at high fractions of their Eddington luminosities could blow away some of their obscuring, dusty-tori via increased radiative pressure, producing a lower fraction of obscured quasars at higher Eddington ratios. \textit{Panel (b)}: Schematic of galaxy-scale gas obscuration \citep{buch17mstar-nh, pann09extinction}. Empirical relationships between $N_{\text{H}}$/covering fraction and galaxy stellar mass have been presented in which less massive galaxies host less obscuring gas than their more massive counterparts. We study these scenarios as possible causes for the mass difference seen in clustering measurements of mid-IR selcted quasars.} \label{fig:schematic}
\end{figure}

However, it is unclear if modeling obscuration as an evolutionary stage is the only way to be able to reproduce the difference in average dark matter halo mass ($M_{\text{halo}}$) between obscured and unobscured quasars. Although the simplest iterations of the unified model are inconsistent with clustering measurements, the dusty torus clearly plays an important role in quasar obscuration \citep[e.g.,][]{netz15unification,ricc17edd,hick18obsc-review}. \citet{ricc17edd} showed that radiative feedback from an AGN could allow for the expulsion of nuclear obscuring dust, thus reducing the number of obscured lines-of-sight between the quasar and observer. Since the amount of radiation pressure exerted on the torus is dependent on the quasar's Eddington ratio ($\lambda_{\text{Edd}}$), it is possible that $\lambda_{\text{Edd}}$ is driving quasar obscuration. The top panel of Figure \ref{fig:schematic} shows a schematic of  ``radiation-regulated unification,'' in which $\lambda _{\text{Edd}}$ determines how much of the quasar is covered by nuclear gas and dust.

It is also possible that quasar obscuration could be taking place in regions outside of the host galaxy's nucleus, but is not associated with a specific galaxy's evolutionary stage. \citet{buch17mstar-nh} analyzed the X-ray afterglows of extragalactic long-duration ($>$ 2s) Gamma ray bursts (LGRBs) to derive host galaxy gas column densities. From this, they determined a relationship between the stellar mass of a host galaxy and its gas column density in which more massive galaxies have larger average $N_{\text{H}}$, thus more of a probability of obscuring a central quasar. \citet{pann09extinction} and \citet{whit17obscSF} found a similar dependence on the fraction of obscured star formation on host galaxy stellar mass. Since star formation in massive galaxies is being obscured by interstellar gas and dust, it may be expected that a central quasar would also be obscured. A schematic of galaxy-scale obscuration is shown in the bottom panel of Figure \ref{fig:schematic}.

In this work, we test these simple models of radiation-regulated unification and galaxy-scale obscuration to determine if they can generate populations of simulated quasars that are consistent with observations of mid-IR selected quasars. We also probe the effect that a luminosity cut that is representative of the limits of \textit{WISE} has on the $M_{\text{halo}}$ of our simulated obscured and unobscured quasar populations.

\begin{table*}[]
    \centering
    \caption{Definitions of terms used throughout this work.}
    \begin{tabular}{l l }
    \hline \hline
    Term & Definition \\
    \hline
    Obscured quasar & Quasar that is shielded by LOS $N_{\text{H}} \gtrsim 10^{22} \ \text{cm}^{-2}$ (e.g., \citealt{usma14obscuration, hick18obsc-review}). \\
    Covering fraction ($f_{\text{cov}}$) & Probability of an observer having an obscured LOS to a quasar based on the physical distribution \\
     &  of obscuring material. (e.g., \citealt{ricc17edd}) \\
    Obscured fraction ($f_{\text{obsc}}$) & Fraction of quasars in a given population that are obscured.\\
    \hline
    \end{tabular}
    \label{tab:definitions}
\end{table*}

Definitions to frequently used terms are given in Table \ref{tab:definitions}. We adopt a cosmology of $H_{0} = 70.2 \ \textrm{km} \textrm{s}^{-1} \textrm{Mpc}^{-1}$,
$\Omega _{M} = \Omega _{CDM} + \Omega _{b} = 0.229+0.046 = 0.275$, and
$\Omega _{\Lambda} = 0.725$ \citep{koma11WMAP}.

\section{The Models}
In this section we describe how we construct our simple models of quasar obscuration based on known halo mass and $\lambda _{\text{Edd}}$ distributions, as well as empirical relationships between obscuring fraction and $\lambda _{\text{Edd}}$ and obscuring fraction and host  galaxy  stellar  mass. 
\subsection{Generating the Quasar Sample}
We begin by generating a model population of 10 million dark matter halos randomly and uniformly distributed in logarithmic space in the mass range,  $10^{10} \ M_{\odot} <M_{\text{halo}}<10^{16} \ M_{\odot}$. Each of these sample halos was assigned a weight using the $z = 1$ halo mass function (HMF) detailed in \citet{tink10bias} so that each halo's contribution to the total average is proportional to the space density of halos of that mass.  We used a \textsc{camb} \citep{lewi00camb} generated matter power spectrum to compute the HMF. Weighting our uniformly and randomly distributed sample of host halos by the HMF eliminates shot noise in our simulated data. This is because that even though our rare, high mass halos will have a small contribution to the average host halo mass, they are still equally as numerous in our simulation as their low mass counterparts.

Once we produced a weighted sample of halos across a wide mass range, we calculated the stellar masses of their corresponding galaxies using the $z=1$ halo mass-stellar mass relationship presented in \citet{most10hm-sm}. The distribution of our simulation galaxies in stellar mass-space is consistent with observed $z \sim 1$ stellar mass functions \citep[e.g.,][]{pere08SMF, behr10SMHM}. We then calculated the masses of the central black holes of each galaxy with the stellar mass-black hole mass relationship detailed in \citet{haer04ej}. Again, we find that the black hole mass distribution of our simulated quasars is broadly consistent with observed black hole mass functions \citep[e.g.,][]{shank09BHpops, kell12BHMF}. There is intrinsic scatter in both the halo mass-stellar mass and stellar mass-black hole mass relationships, so we included these effects in our models. We adopted an intrinsic scatter of 0.2 dex for the halo mass-stellar mass and 0.3 dex for the stellar mass-black hole mass relationships \citep{haer04ej,most10hm-sm}.

We generated a separate sample of 10 million Eddington ratios that are randomly and uniformly distributed in logarithmic space in the range, $-4 < \log \lambda _{\text{Edd}} < 1$. Just as we assigned weights to each dark matter halo based on the HMF, we also assigned weights to each $\lambda _{\text{Edd}}$ that correspond to the double power law-$\lambda _{\text{Edd}}$ distribution presented in \citet{jone19galform} to limit the contribution of rare, high Eddington systems to the overall distribution. The overall probability of a halo of a given mass containing a quasar accreting at a particular $\lambda _{\text{Edd}}$ is the product of the HMF and the $\lambda _{\text{Edd}}$ distribution function. 

Although we can use this treatment to generate quasars of all luminosities, observational surveys are limited by their capabilities to detect faint sources. Our model therefore needs to include a lower luminosity limit so we can match our quasar distributions to observations. We first calculate the bolometric luminosities for all of our generated quasars. Since we are interested in mid-IR selected quasars, we implement a luminosity threshold that is representative of the detection limits of \textit{WISE}. Bolometric luminosities for \textit{WISE}-selected quasars at $z=1$ are typically greater than $10^{46} \ \textrm{erg s}^{-1}$ \citep[e.g.,][]{hick07bootes,asse13wise-quasars}, so we impose a luminosity limit of $10^{45.8} \ \textrm{erg s}^{-1}$ following \citet{dipo17model} unless otherwise stated.

\subsection{Identifying Obscured Sources} \label{ssec: identify}
Creating obscured and unobscured populations of quasars from the simulated sample requires us to adopt a model that parameterizes obscuration as a function of one of the physical properties of either the quasars or their host galaxies. Broad band observations of quasars can tell us whether or not a given source is obscured \citep[e.g.,][]{merl14AGNobsc}, but they do not necessarily yield information on what scale the light emitted from the quasar is being absorbed. We first assume that our quasars are being obscured by their dusty tori, and adopt the radiation-regulated unification model in which obscuration is parameterized by the $\lambda _{\text{Edd}}$ of our quasars \citep{ricc17edd}. For galaxy-scale interstellar material, we parameterize obscuration as a function of host galaxy stellar mass \citep{pann09extinction, buch17mstar-nh, whit17obscSF}. We also present a model that allows our simulated quasars to be obscured by both,  their tori and the interstellar material in their host galaxies. In what follows, we describe the details of each model.
\subsubsection{Radiation-regulated Unification Model} \label{sssec: ricci}
\citet{ricc17edd} presented the relationship between the covering fractions of AGN and their $\lambda _{\text{Edd}}$. This relationship was derived from a multi-wavelength study of 836 AGN identified by the \textit{Swift}/Bat X-ray survey 
\citep[e.g.,][]{gehr04Swift,bart05BAT,krimm13xraytrans,baum13BATallsky}. We used the observed relationship shown in in Figure 4 of \citet{ricc17edd} to model a population of obscured and unobscured AGN where the obscured fraction depends on $\lambda _{\text{Edd}}$.  We chose to use this relationship over the one detailed in Figure 1 of \citet{ricc17edd} to account for the existence of Compton-thick material that might obscure the most highly accreting quasars. Although this was originally presented for AGN at $z \sim 0.1$, we expect it to hold for our model at $z=1$. Observations of high redshift quasars have shown that there is not much evolution over cosmic time on the $\lesssim 1$ parcsec scale at which radiation-regulated feedback would be significant \citep[e.g.,][]{fan06highzquasar, luss16xrayUV}.
\begin{figure}
\begin{center}
\includegraphics[width=0.5\textwidth]{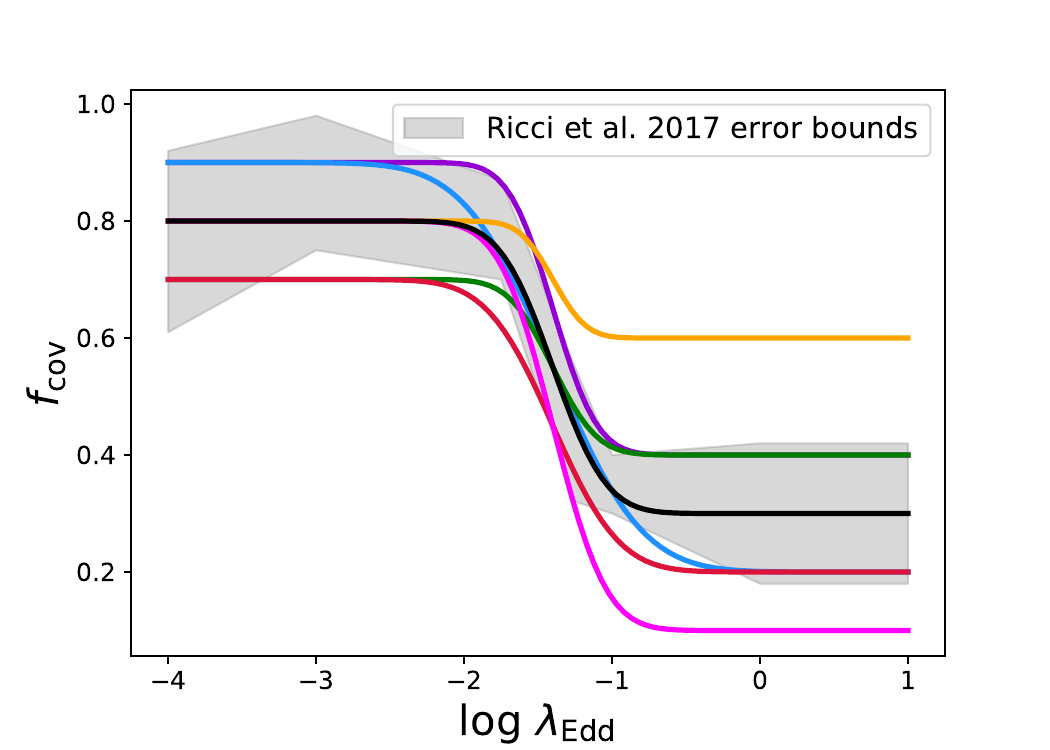}
\end{center}
\caption{Relationship between covering fraction and $\lambda _{\text{Edd}}$ for the radiation-regulated unification model presented in \citet{ricc17edd}. The error bounds from Figure 4a in \citet{ricc17edd} are shown in grey. We modeled this relationship as error functions that spanned the parameter space occupied by the error bounds in \citet{ricc17edd}. We also included 2 model fits that fall outside of the \citet{ricc17edd} error bounds to account for the uncertainty on the Compton-thick fraction.} \label{fig:ricci}
\end{figure}

The data bins used in \citet{ricc17edd} to average covering fractions at a given $\lambda _{\text{Edd}}$ were broad, so we made this relationship more continuous over a range of $\lambda _{\text{Edd}}$ by fitting a series of error functions to the original data, as seen in Figure \ref{fig:ricci}. Each fit corresponds to varying the minimum $f_{\text{cov}}$ for high accreting quasars (the covering fractions in the Compton-thick regime are not well constrained). The grey, shaded region in Figure \ref{fig:ricci} represents the errors on the $ f_{\text{cov}}- \log \lambda _{\text{Edd}}$ relationship shown in Figure 4 of \citet{ricc17edd}. Using these $f_{\text{cov}}- \log \lambda _{\text{Edd}}$ relationships, we then randomly assigned the quasars to obscured and unobscured populations. We do this by assigning each quasar a random number between zero and one. If this number is less than or equal to  $f_{\text{cov}}$ at a quasar's $\lambda_{\text{Edd}}$, then it is classified as obscured. Otherwise, it is classified as unobscured.

We also note that for some populations of quasars at higher redshifts that it is possible for the accretion disk to have a ``slim disk'' geometry in which the accretion disk is puffed up for the quasars that are accreting at high-$\lambda _{\text{Edd}}$ \citep[e.g.,][]{fran02accretion,leig04slimdisk,luo15slimdisk}. The thin disk is geometrically thin and optically thick, so in principle this could contribute to obscuration in addition to the dusty-torus. As for the $f_{\text{cov}}-\lambda _{\text{Edd}}$ relationships in Figure \ref{fig:ricci}, this effect would increase $f_{\text{cov}}$ again at high-$\lambda _{\text{Edd}}$. We consider the effect of slim-accretion disk geometries below, but this scenario is more applicable for luminous quasars at $z \sim 2$ than the population we are simulating \citep[e.g.,][]{netz14ADevolution}.

\subsubsection{Galaxy-Scale Obscuration} \label{ssec:galaxy-sclae}
The radiation-regulated unification model assumes that the quasars are being obscured by the parsec-scale dusty torus, and that the $\lambda _{\text{Edd}}$ of the quasar could change the covering fraction of the torus. However, toroidal dust is not the only obscuring material in front of the quasar along the LOS of an observer. Interstellar gas and dust within a galaxy could have the ability to obscure a quasar at the galactic center \citep[e.g.,][]{hick18obsc-review}. 
\begin{figure} 
\begin{center}
\includegraphics[width=0.5\textwidth]{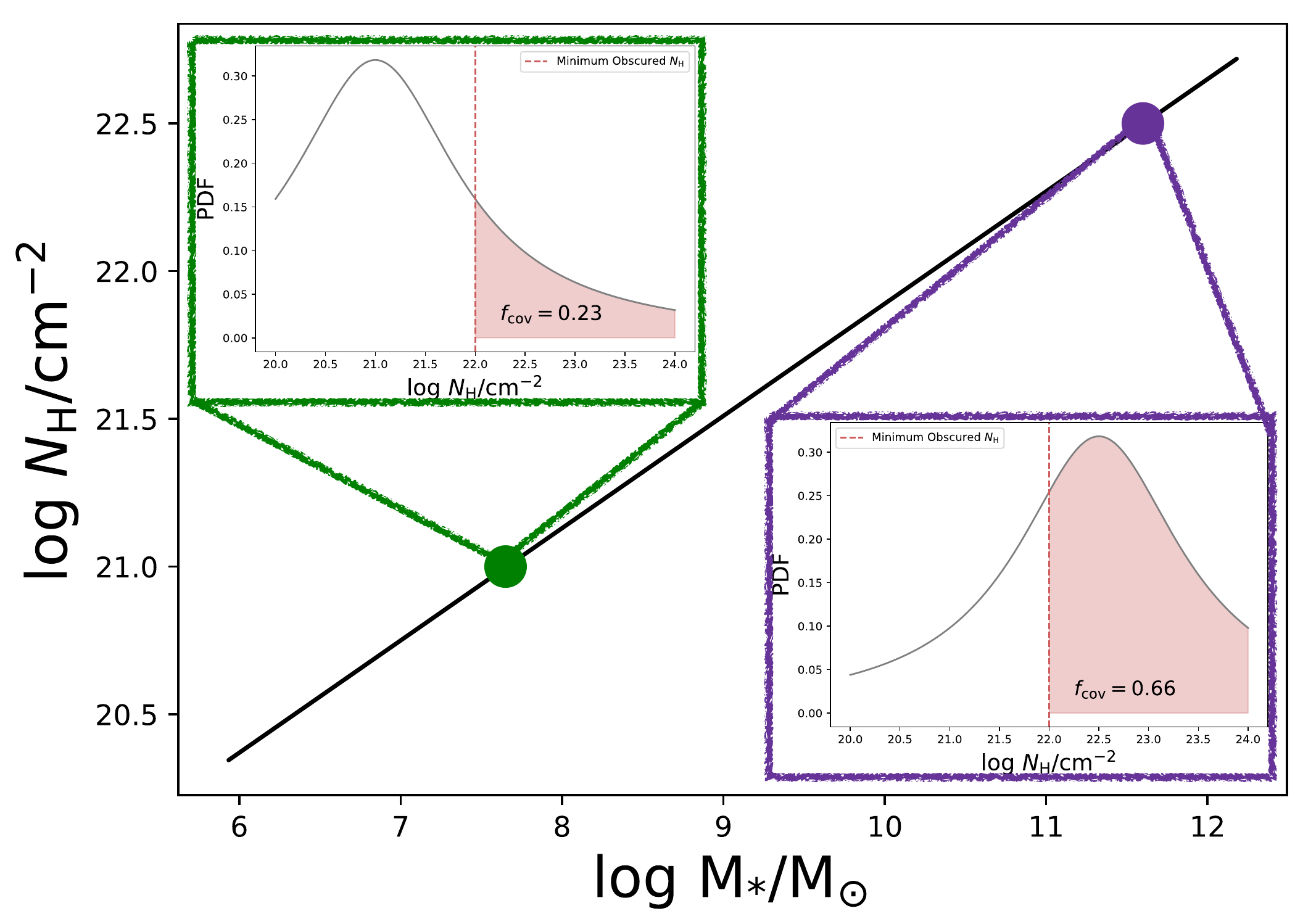}
\end{center}
\caption{Schematic of how covering fractions are calculated from assigned $N_{\text{H}}$ values. Each galaxy is assigned a mean $N_{\text{H}}$ based on its $M_{*}$. Each galaxy's mean $N_{\text{H}}$ is then used as the mean of a column density probability density function (PDF) that is then integrated on the interval $10^{22} < N_{\text{H}}/ \text{cm}^{-2} < \infty$ to determine covering fraction at each given $M_{*}$.} \label{fig:nh_mstar_buch_schem}
\end{figure}
\begin{figure} 
\begin{center}
\includegraphics[width=0.5\textwidth]{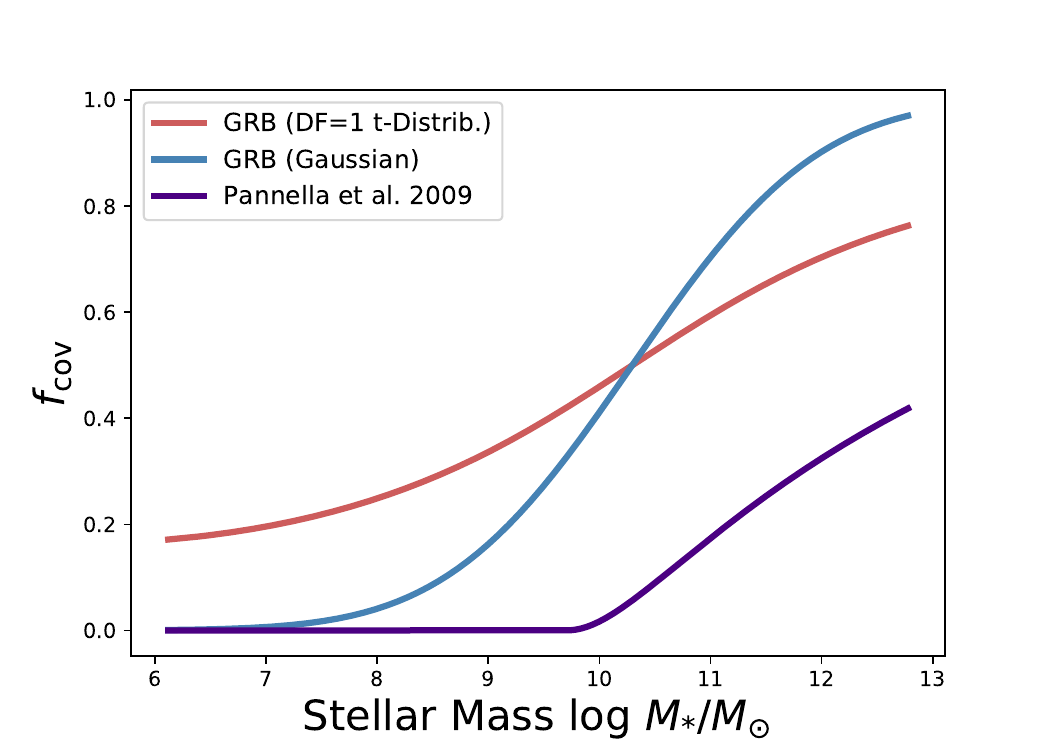}
\end{center}
\caption{Relationships between covering fraction and host galaxy stellar mass. Using the mean $N_{\text{H}}$ at a given stellar mass, we calculated covering fractions as detailed in Section \ref{ssec:galaxy-sclae}. These relationships are used to produce model populations of obscured and unobscured quasars based on obscuration by galaxy scale gas.} \label{fig:nh_mstar_buch}
\end{figure}
\citet{buch17mstar-nh} measured the attenuation of X-ray afterglows from extragalactic LGRBs to derive an empirical relationship between the mean column densities of gas in galaxes and their stellar masses.  This relationship shows that more massive galaxies contain deeper obscuring columns of gas. Knowing the $N_{\text{H}}$ of gas in galaxies can allow us to determine the likelihood of obscuration for a given quasar. \citet{buch17AGNobscuration} used the $\log \ N_{\text{H}} - \log \ M_{*}$ relationship derived in \citet{buch17mstar-nh} to construct a simple model of obscuring covering fractions for AGN. Here, we study this model to determine if it is capable of recreating clustering measurements of mid-IR selected quasars.

We start by using the GRB-derived $\log \ N_{\text{H}} - \log \ M_{*}$ relationship from \citet{buch17mstar-nh} to assign each of our simulated galaxies a mean $N_{\text{H}}$. We assume that the assigned $N_{\text{H}}$ is the mean of a column density probability distribution. Here, we use a Gaussian probability density function with $\sigma = 0.5$, as well as a 1 degree-of-freedom Student's $t$-distribution. The Student's $t$-distribution acts as a proxy for the broader, less peaked SingleEllipse model detailed in \citet{buch17mstar-nh} since the two models have a similar analytic form (private communication; J. Buchner). X-ray selected AGN are typically detected as obscured when $N_{\text{H}}> 10^{22}\  \text{cm}^{-2}$ \citep[e.g.][]{pred95ISM-NH,burt16seyfert_obsc,schn16BLR-extinct}, and this generally corresponds to the $N_{\text{H}}$ of mid-IR selected quasars \citep[e.g.,][]{hick07bootes, usma14obscuration}. We convert the mean column densities from the $\log N_{\text{H}} - \log M_{*}$ relationships to effective covering fractions by integrating each of the $N_{\text{H}}$ probability density functions on the interval $10^{22} < N_{\text{H}}/ \text{cm}^{-2} < \infty$, as depicted in Figure \ref{fig:nh_mstar_buch_schem}. The covering fraction-stellar mass relationships derived using the Gaussian and Student's $t$-distributions are shown as the blue and red curves in Figure \ref{fig:nh_mstar_buch}, respectively.

In addition to the GRB X-ray afterglow attenuation-derived models described above, we also calculated a $f_{\text{cov}}- \log M_{*}$ relationship based on the galaxy mass dependence of the fraction obscured star formation in galaxies presented in \citet{whit17obscSF}. The simple assumption here is that the material obscuring star formation in these galaxies will similarly obscure quasar activity. \citet{pann09extinction} presented a relationship between ultraviolet (UV) attenuation and stellar mass. We utilized this relationship to derive LOS column densities as a function of stellar mass since it is unclear how the fraction of obscured star formation in a galaxy relates to the physical dust distribution. \citet{whit17obscSF} showed that the obscured star formation fractions derived from the \citet{pann09extinction} relationship were consistent with what they calculated from IR and UV star formation rates.

We convert the \citet{pann09extinction} UV attenuation-stellar mass relationship to a column density-stellar mass relationship by assuming $R(V) = 3.1$ (Milky Way) extinction curve \citep[e.g.,][]{fitz99extinction, drai03dust}. At $1500 \ \text{\AA}$, this corresponds to  $A_{1500}/N_{\text{H}} = 1.6 \times 10^{-21} \frac{\text{cm}^{2} \text{mag}}{\text{H}}$. We note that the UV attenuation-stellar mass relationship in \citet{pann09extinction} is fitted over a much smaller stellar mass range than included in our simulated sample. However, once we enact a luminosity threshold, only $\sim$12\% of our sources fall outside the \citet{pann09extinction} stellar mass range, and of those sources $\sim$88\% fall within 0.3 dex of the fitted mass range, so we are confident in the extrapolation of this relationship. We then compute a $f_{\text{cov}}-M_{*}$ relationship using the same methodology as done with the models derived from the attenuation of GRB X-ray afterglows.

\section{Results}
Our models need to be able to recover the following observational constraints: (1) the host $M_{\text{halo}}$ for our simulated obscured and unobscured quasars, as well as (2) the fraction of obscured quasars. The measured average host $M_{\text{halo}}$ of obscured and unobscured quasars are $\log M_{\text{halo}}/M_{\odot} = 12.94 ^{+ 0.10} _{- 0.11}$ and $\log M_{\text{halo}}/M_{\odot} = 12.49 ^{+ 0.08} _{- 0.08}$, respectively \citep[e.g.,][]{dipo17qsoclust}. The range of observed obscured fractions for luminous quasars is roughly between 30\% \citep{trei08fobsc} and 65\% \citep{poll08obscuration}, with significant uncertainty on the heavily obscured (Compton-thick) population \citep[e.g.,][]{dipo16torus, yan19obscWISE}. We note that we adopt such a broad observed obscured fraction to reflect the uncertainty due to the difficulty of detecting heavily-obscured AGN. This is a conservative estimate that provides a broad parameter space in which our models could be potentially viable.

\subsection{Radiation-regulated Unification Model}
As seen above in Figure \ref{fig:ricci}, we modeled radiation-regulated unification as a series of error functions within the $f_{\text{cov}}-\lambda _{\text{Edd}}$ parameter space covered by the error bounds of the relationship shown in Figure 4a of \citet{ricc17edd}. We also included two parameterizations that were well above and below the error bounds to account for the uncertainty in the Compton-thick fraction of quasars. We calculated $f_{\text{obsc}}$ for each of these models and found that $f_{\text{obsc}}$ is roughly equal to the value of $f_{\text{cov}}$ at high-$\lambda_{\text{Edd}}$. This is because the luminosity cut pushes the mean of the underlying Eddington ratio distribution to be $\log \langle \lambda _{\text{Edd}} \rangle \approx 0$. Since there is little dynamic range in $f_{\text{cov}}$ at high-$\lambda _{\text{Edd}}$, $f_{\text{obsc}}$ becomes the assigned high-$\lambda _{\text{Edd}}$ $f_{\text{cov}}$ value. The implication of this on our simulated populations is that only the three parameterizations with high-$\lambda _{\text{Edd}}$ $f_{\text{cov}} \geq 0.3$ satisfy $f_{\text{obsc}}$ constraints. In what follows, we focus on the model fit (shown in black in Figure \ref{fig:ricci}) that is the mean of the \citet{ricc17edd} error bounds. We use this relationship since it produced a population of quasars whose $f_{\text{obsc}}$ falls on the edge of the observed obscured fraction range, as well as that it best represents the results presented in \citet{ricc17edd}.
\begin{figure} 
\begin{center}
\epsscale{1.25}
\includegraphics[width=0.5\textwidth]{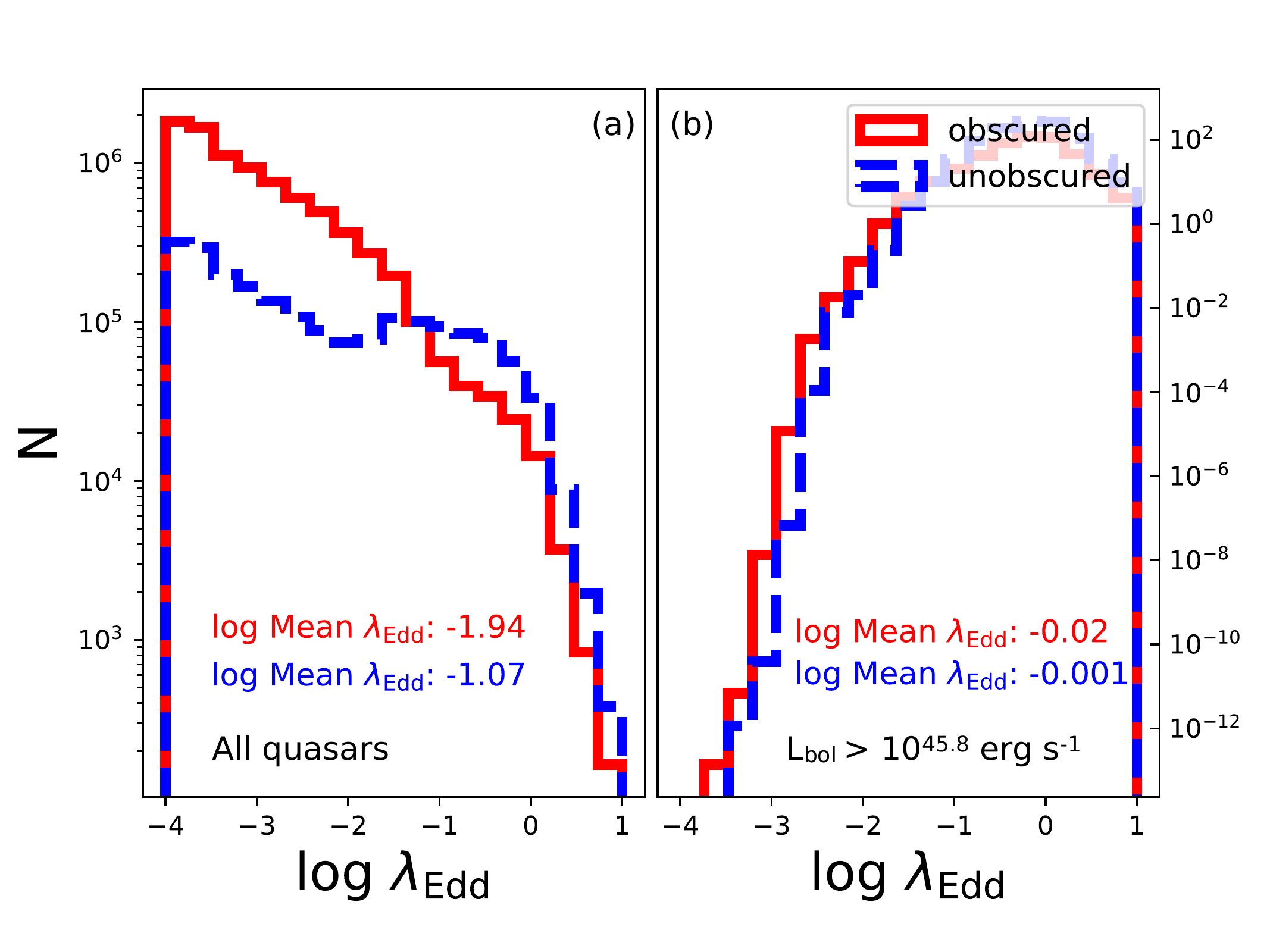}
\end{center}
\caption{\textit{Panel (a)}: The full weighted distributions of $\lambda _{\text{Edd}}$ for our simulated quasars generated using the mean of the \citet{ricc17edd} error bounds on $f_{\text{cov}} - \log \lambda _{\text{Edd}}$ (black curve in Figure \ref{fig:ricci}). Obscured quasars are shown in red bins, and unobscured quasars in blue. There is an intrinsic difference in the $\lambda _{\text{Edd}}$ distributions between the obscured and unobscured populations of quasars due to the fact that the chosen $f_{\text{cov}}-\log \lambda _{\text{Edd}}$ relationship preferentially obscures low-$\lambda _{\text{Edd}}$ quasars. \textit{Panel (b)}: The distribution of $\lambda _{\text{Edd}}$ after a luminosity cut of $10^{45.8} \text{erg} \ \text{s}^{-1}$, corresponding to \textit{WISE}-selected quasars. \citep[e.g.,][]{dipo17model}. The luminosity cut causes our model to exclude the low-$\lambda _{\text{Edd}}$ end of our initial distributions, thus pushing our populations to become increasingly} similar.\label{fig:dipompeo_ledd_dist}
\end{figure}
\begin{figure} 
\begin{center}
\epsscale{1.25}
\includegraphics[ height=6cm,width=0.5\textwidth]{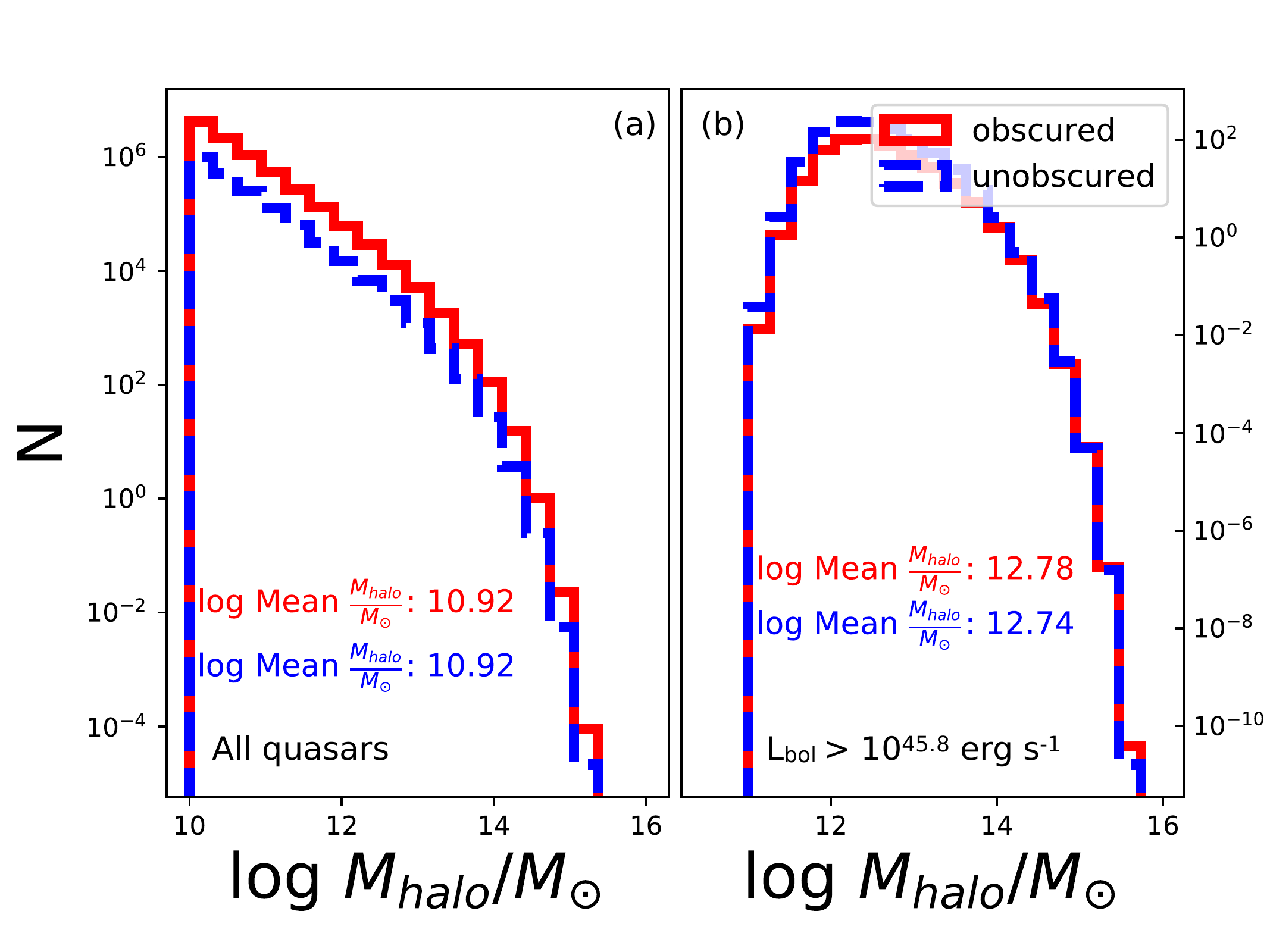}
\end{center}
\caption{\textit{Panel (a)}: The full weighted distributions of host $M_{\text{halo}}$ for our simulated quasars generated using the mean of the \citet{ricc17edd} error bounds on $f_{\text{cov}} - \log \lambda _{\text{Edd}}$ (black curve in Figure \ref{fig:ricci}). Obscured quasars are shown in red bins, and unobscured quasars in blue. \textit{Panel (b)}: The distributions of host $M_{\text{halo}}$ for our sample quasars after a luminosity cut of $10^{45.8} \text{erg} \ \text{s}^{-1}$, corresponding to \textit{WISE}-selected quasars. \citep[e.g.,][]{dipo17model}. The obscured and unobscured populations have the same mean $M_{\text{halo}}$ before the luminosity cut, and only a negligible post-cut difference that falls outside of our observational constraint on mean $M_{\text{halo}}$.\label{fig:dipompeo_dist}}
\end{figure}

For this $f_{\text{cov}}- \log \lambda _{\text{Edd}}$ relationship, we examined the $\lambda _{\text{Edd}}$ and host $M_{\text{halo}}$ distributions for the generated obscured and unobscured populations of quasars. Figure \ref{fig:dipompeo_ledd_dist} presents the full $\lambda _{\text{Edd}}$ distribution for our simulated quasars as well as the distribution after a luminosity cut of $10^{45.8} \ \text{erg} \ \text{s}^{-1}$ has been applied \citep[e.g.,][]{dipo17model}. Initially, there is an intrinsic difference between the shapes of the obscured and unobscured $\lambda _{\text{Edd}}$ distributions. As expected, the unobscured population has a higher mean $\lambda _{\text{Edd}}$ than its obscured counterpart due to the fact that the shape of the $f_{\text{cov}}- \log \lambda _{\text{Edd}}$ distribution dictates that low-$\lambda _{\text{Edd}}$ quasars have a higher probability of being obscured. However, applying a lower luminosity limit causes us to lose the low-$\lambda _{\text{Edd}}$ end where the two populations are the most distinct from one another. This effectively makes the mean $\lambda _{\text{Edd}}$ identical for the populations of obscured and unobscured quasars. 

Figure \ref{fig:dipompeo_dist} shows the corresponding host $M_{\text{halo}}$ distributions. In this model, obscuration is independent of host $M_{\text{halo}}$, so our distributions for the full populations of simulated obscured and unobscured quasars are identical. The initial $f_{\text{obsc}}$ for the full sample is 78\%, dropping to 29\% for the luminosity cut; thus removing a significant number of our obscured quasars that reside in low-mass dark matter halos. The removal of obscured quasars in low mass halos results in a small difference between the average host $M_{\text{halo}}$ for the obscured and unobscured populations, but it is still well outside of our observational constraints. 

\begin{figure} 
\begin{center}
\includegraphics[width=0.5\textwidth]{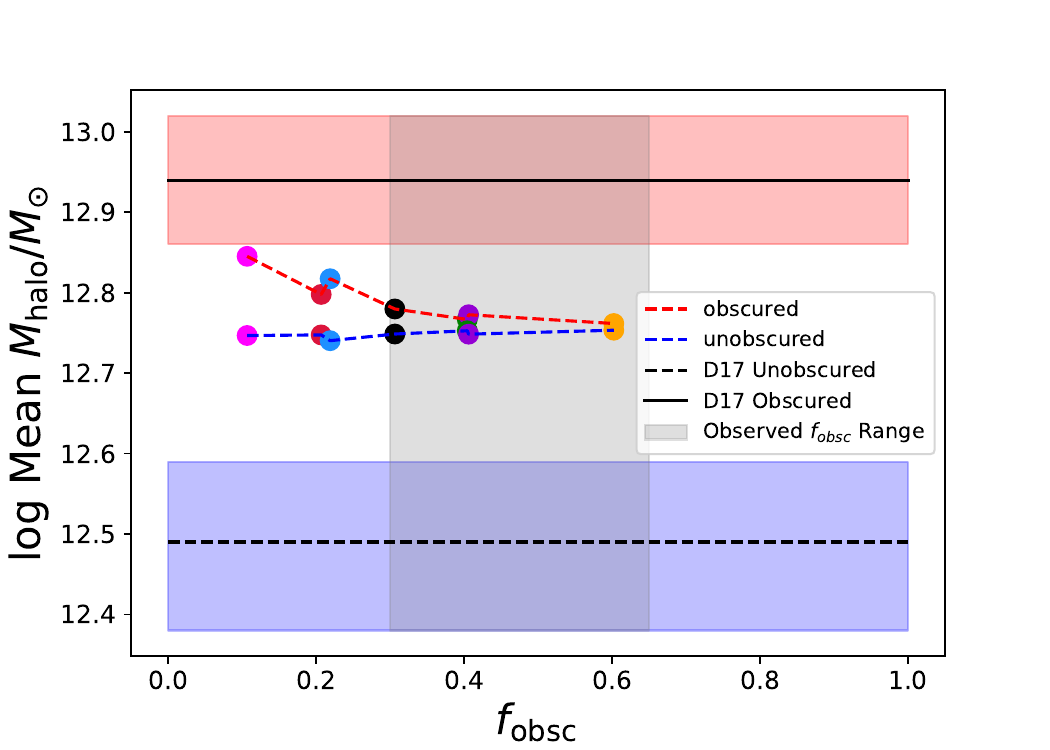}
\end{center}
\caption{The calculated mean halo masses for simulated quasar populations generated using different error function fits to the $f_{\text{cov}} - \log \lambda _{\text{Edd}}$ relationship as seen in Figure \ref{fig:ricci}. The solid, black line shows the measured mean halo mass of a population of observed obscured quasars \citep[e.g.,][]{dipo17qsoclust}, where the red, shaded region shows the error on that measurement. This is also the case for the black, dashed line, and the blue, shaded region, but for the unobscured population studied in \citet{dipo17qsoclust}. Each point corresponds to the populations generated using the model of the same color in Figure \ref{fig:ricci}. The points connected by the red (blue), dashed line are the average obscured (unobscured) host halo masses. The radiation-regulated unification model is unable to recover both the disparity in $M_{\text{halo}}$ between obscured and unobscured quasars and an obscured fraction that falls within the range of observations.} \label{fig:med_halo_vs_slope}
\end{figure}

We next carry out this analysis for all of the parameterizations of our radiation-regulated unification model, as seen in Figure \ref{fig:ricci}. Just as we calculated the fraction of obscured quasars for each $f_{\text{cov}} - \log \lambda _{\text{Edd}}$ relationship, we also calculated mean host $M_{\text{halo}}$ for the generated populations of obscured and unobscured quasars. These are presented in Figure \ref{fig:med_halo_vs_slope}.  The red and blue shaded regions show the uncertainty for the measured mean $M_{\text{halo}}$ for mid-IR selected obscured and unobscured quasars, respectively \citep{dipo17qsoclust}. We find that as we increase the covering fraction at high-$\lambda _{\text{Edd}}$, the mean $M_{\text{halo}}$ for obscured and unobscured quasars become increasingly similar. Increasing the covering factor for high-$\lambda _{\text{Edd}}$ sources at a given luminosity threshold allows for more low-mass, high-$\lambda _{\text{Edd}}$ quasars to be classified as obscured. For a luminosity cut of $10^{45.8} \ \text{erg} \ \text{s}^{-1}$, there is nowhere in this parameter space that satisfies both the mass difference and obscured fraction observational constraints.

\begin{figure} 
\begin{center}
\includegraphics[width=0.5\textwidth]{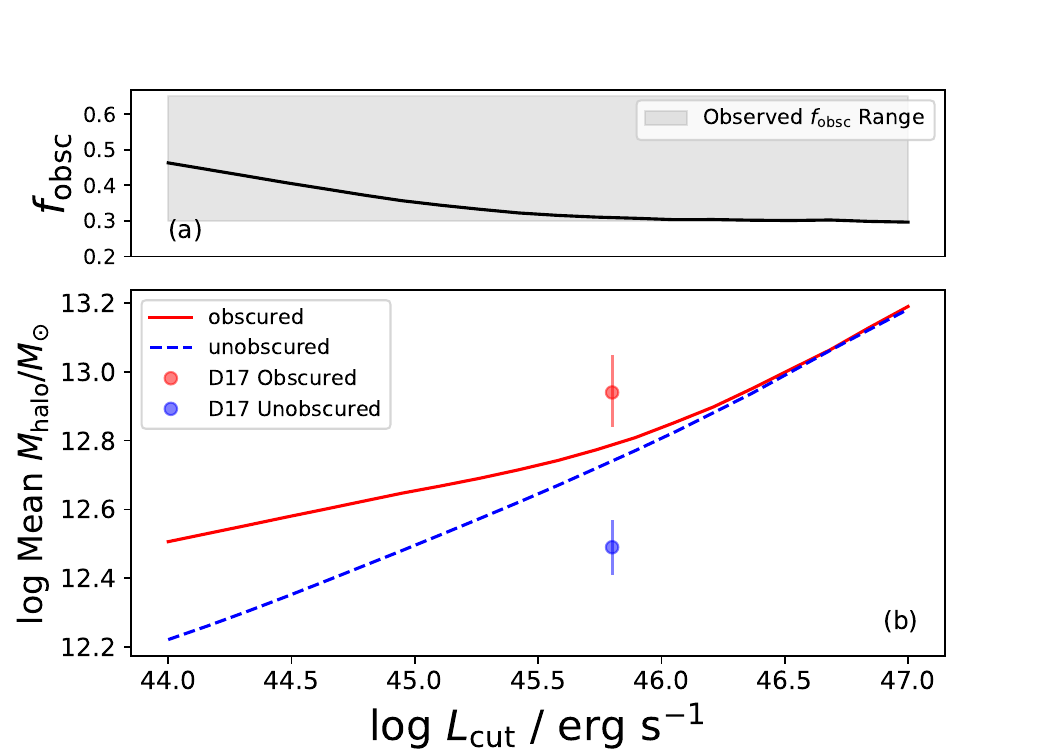}
\end{center}
\caption{\textit{Panel (a)}: The obscured fraction of our sample population modeled from the yellow curve in Figure \ref{fig:ricci} as a function of the luminosity threshold for the radiation-regulated unification model. At every luminosity limit, the obscured fraction resides within the range of observed obscured fractions. \textit{Panel (b)}: The relationship between the weighted mean $M_{\text{halo}}$ of the distribution as a function of the luminosity threshold. The obscured sample is depicted by the solid, red line, and the unobscured is depicted by the dashed, blue line. It is apparent that the choice in luminosity limit affects the disparity between the mean $M_{\text{halo}}$ for obscured and unobscured quasars, but it does not reproduce observations.} \label{fig:avg_halo_vs_lcut}
\end{figure}

Following \citet{dipo17model}, we probed the effect the luminosity cut had on our simulated quasar populations. For the population of quasars generated from the minimum $f_{\text{cov}} = 0.3$ model, we find that as the luminosity cut increases, the mean $M_{\text{halo}}$ for the obscured and unobscured populations converge. Increasing the minimum detectable luminosity effectively pushes our sample to be comprised of quasars that are either accreting at higher $\lambda _{\text{Edd}}$ or residing in higher mass dark matter halos. As also shown in Figures \ref{fig:dipompeo_ledd_dist} and \ref{fig:dipompeo_dist}, increasing the lower luminosity limit excludes the low-Eddington end of our quasar populations where their $\lambda _{\text{Edd}}$ distributions are most distinct from one another. Our model is able to produce a $\sim 0.3$ dex difference in $M_{\text{halo}}$ for obscured and unobscured quasars at low luminosity cuts (around $10^{44} \ \textrm{erg} \ \textrm{s}^{-1}$), which is still even smaller than the observed difference shown in \citet{dipo17qsoclust}. This is shown in Figure \ref{fig:avg_halo_vs_lcut}. Overall, the difference between simulated $M_{\text{halo}}$ for our obscured and unobscured populations fall significantly below observations. 

As mentioned earlier at the end of Section \ref{sssec: ricci}, we also considered the effect of obscuration due to a slim accretion disk at high-$\lambda _{\text{Edd}}$. We did this by implementing a linear increase of $f_{\text{cov}}$ starting at $\log \lambda _{\text{Edd}}=0$ such that a quasar with $\log \lambda _{\text{Edd}} = 1$ has a covering fraction of 1. We found that implementing a slim accretion disk to the model with minimum $f_{\text{cov}}=0.3$ mildly decreased the average $M_{\text{halo}}$ for our obscured population of quasars, making it identical to the average unobscured quasar dark matter halo mass. The average $M_{\text{halo}}$ for the simulated obscured and unobscured quasars are both $\log M_{\text{halo}}/M_{\odot} = 12.75$. This, in addition to the fact slim accretion disks are also more often found in luminous, $z \sim 2$ AGN rather than in the $z \sim 1$ quasar populations we are modeling \citep[e.g.,][]{netz14ADevolution}, shows us that this model is not viable for recreating mid-IR quasar clustering measurements.

\subsection{Galaxy-scale Gas Obscuration}
Here, we conduct a similar analysis as for the radiation-regulated unification model, instead assuming the obscurer is galaxy-scale gas, to determine if this model could satisfy observational constraints. 

\begin{figure} 
\begin{center}
\includegraphics[width=0.5\textwidth]{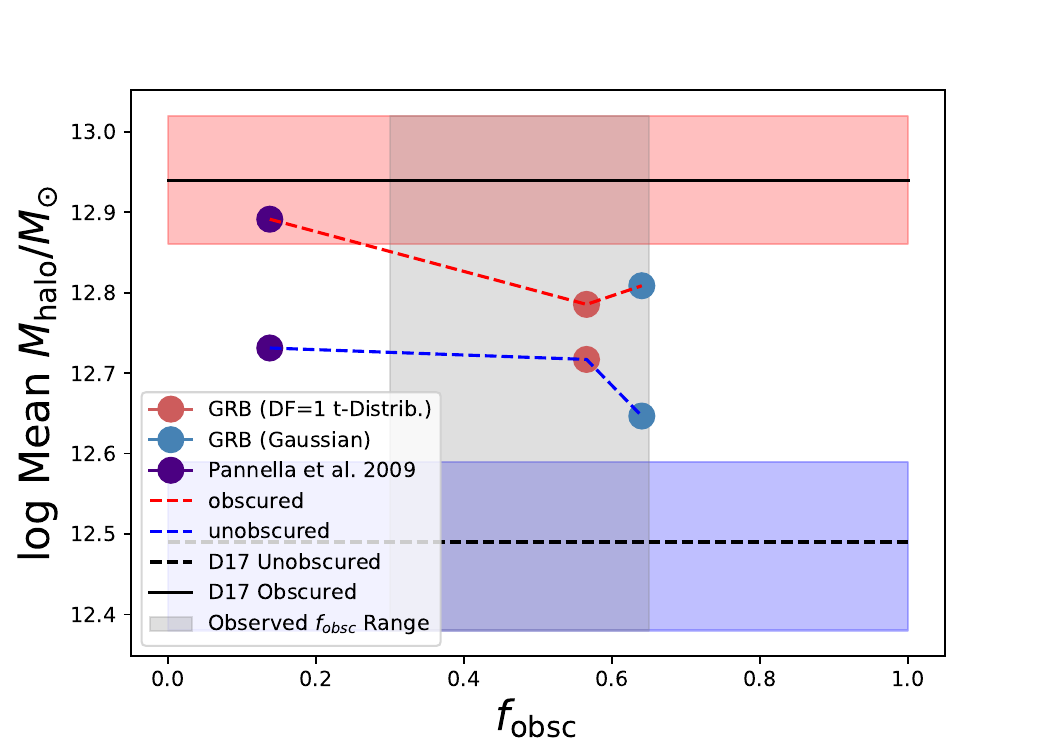}
\end{center}
\caption{The calculated mean $M_{\text{halo}}$ for simulated quasar populations generated using $f_{\text{cov}}-\log M_{*}$} relationships as seen in Figure \ref{fig:nh_mstar_buch}. The red (blue), dashed line connects the average host halo mass for the obscured (unobscured) population generated from the $f_{\text{cov}}-\log M_{*}$ of the same color from Figure \ref{fig:nh_mstar_buch}. The solid, black line shows the measured mean $M_{\text{halo}}$ of a population of observed obscured quasars \citep[e.g.,][]{dipo17qsoclust}, where the red, shaded regions show the errors on that measurement. This is also the case for the black, dashed line, and the blue, shaded region, but for the unobscured population studied in \citet{dipo17qsoclust}. Although these models are able to drive small differences in average $M_{\text{halo}}$ for obscured and unobscured quasars, they do not satisfy observational constraints. \label{fig:mhalo_buch}
\end{figure}

As before, we calculated obscured fractions and mean $M_{\text{halo}}$ values for the populations of quasars that were generated using the various $f_{\text{cov}}-\log M_{*}$ relationships shown in Figure \ref{fig:nh_mstar_buch}. The calculated obscured fraction for each $f_{\text{cov}}-\log M_{*}$ relationship is shown as the x-axis of Figure \ref{fig:mhalo_buch}. It is apparent that two of the $f_{\text{cov}}-\log M_{*}$ relationships produced populations of quasars that were more highly obscured than what has been observed since the points for these models fall outside of the grey box that depicts the range of observed obscured fractions. For a quasar to be luminous enough to be detectable using mid-IR color selection, it would have to be rapidly accreting or host a massive black hole. Since we are considering the galaxy stellar mass-dependent model here as well as scaling relationships between $M_{\text{BH}}-M_{*}$, the quasars that would be  detectable in this model will typically reside in galaxies with large stellar masses. Since the $f_{\text{cov}}-\log M_{*}$ relationships in Figure \ref{fig:nh_mstar_buch} state that more massive galaxies have a higher probability of obscuring the central quasar, this results in the populations of quasars generated with the Gaussian \citet{buch17mstar-nh} inspired-model to have a higher obscured fraction than observed. The mean $M_{\text{halo}}$ values are shown as the y-axis in Figure \ref{fig:mhalo_buch}. Much like what occurred in the results of the radiation-regulated unification model, we find that the weighted mean parent $M_{\text{halo}}$ for all of the $f_{\text{cov}}- \log M_{*}$ relationships fall outside of the range of clustering measurements which is shown by the red (blue) shaded region for observed obscured (unobscured) sources. 
\begin{figure} 
\begin{center}
\includegraphics[width=0.5\textwidth]{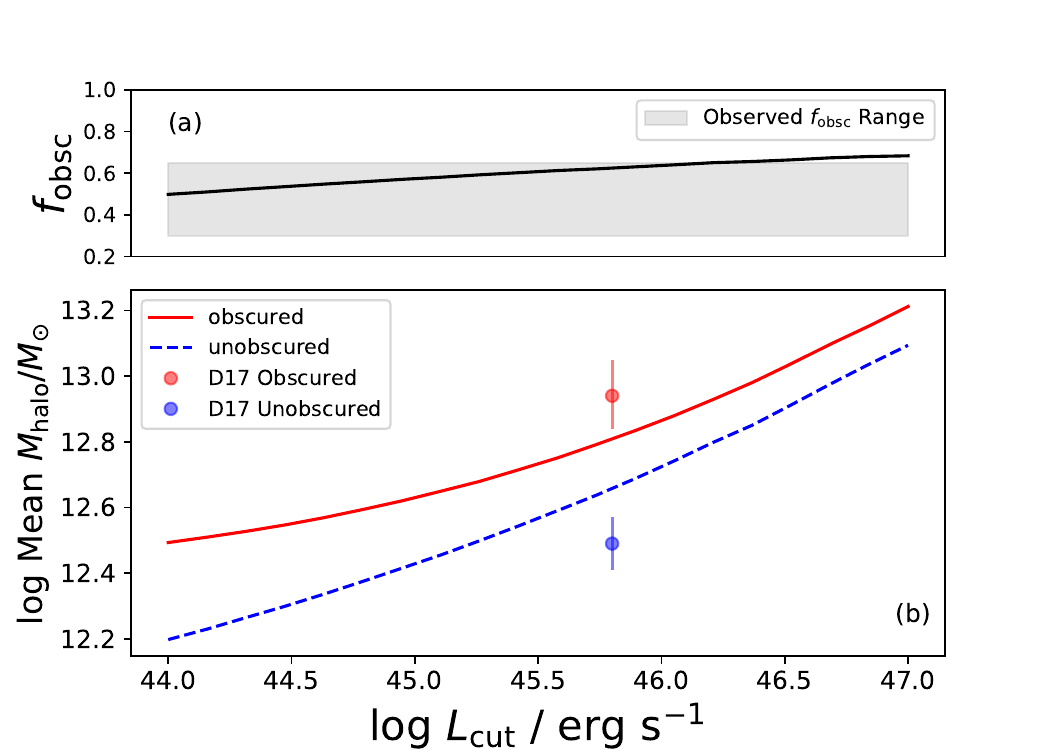}
\end{center}
\caption{\textit{Panel (a)}: The obscured fraction for the populations of quasars generated at each varying luminosity threshold for the observed GRB-derived Gaussian galaxy-scale dust model of obscuration (as shown in blue in Figure \ref{fig:nh_mstar_buch} and thereafter). \textit{Panel (b)}: The relationship between the mean $M_{\text{halo}}$ of our obscured (red, solid curve) and unobscured (blue, dashed curve) populations and luminosity threshold. For this obscuration model, the choice  in  luminosity  limit  minimally affects the disparity between the mean $M_{\text{halo}}$ for obscured and unobscured quasars, but it does not reproduce observations.} \label{fig:avg_halo_vs_lcut_buch}
\end{figure}

 Figure \ref{fig:avg_halo_vs_lcut_buch} shows our GRB-derived Gaussian galaxy-scale gas obscuration model's dependence on luminosity cut. We chose this model because it produced a $f_{\text{obsc}}$ that fell on the edge of observational constraints as well as having a modest difference between $M_{\text{halo}}$ of its obscured and unobscured quasar populations. The intrinsic $M_{\text{halo}}$ distributions created by the galaxy-scale gas models are most different from one another at low-$M_{\text{halo}}$. Once again, a higher luminosity cut results in sampling a region in the original $M_{\text{halo}}$ distribution where the obscured and unobscured distributions are almost indistinguishable. It is clear that modeling a quasar's obscuration as a function of its host galaxy's stellar mass is not sufficient to properly recover clustering measurements of parent $M_{\text{halo}}$ as well as the observed quasar obscured fraction.

\subsection{Combining Nuclear and Galaxy-scale Obscuration}
Both of the models discussed above assume that obscuration is coming from material \textit{either} within the region closest to the quasar or within the interstellar regions of the host galaxy. We next consider that there could be many possible lines-of-sight in which the obscuring material is independently contributed by both, the nuclear-scale torus and the galaxy-scale gas and dust. Here, we adopt a $\log N_{\text{H}}- \log \lambda_{\text{Edd}}$ relationship from \citet{ricc17edd} to assign our quasars nuclear column densities. We then sum the nuclear and the \citet{buch17mstar-nh}-assigned galaxy-scale column densities to obtain a mean LOS column density for each of the sources in our simulated sample. Utilizing the same methodology described above in Section \ref{ssec:galaxy-sclae}, these mean LOS column densities are treated as the mean of a Gaussian PDF that is integrated on the interval $10^{22} <  N_{\text{H}}/\text{cm}^{-2} < \infty$ to obtain a LOS covering fraction for each quasar. We again randomly assign our quasars into obscured and unobscured populations based on their calculated covering fractions.  We find that after applying the luminosity cut, the mean $M_{\text{halo}}$ for the obscured and unobscured populations are  $\log M_{\text{halo}}/ M_{\odot} = 12.79$ and $\log M_{\text{halo}}/ M_{\odot} = 12.65$, respectively, and that $f_{\text{obsc}} = 0.75$. This model overpredicts the number of obscured quasars in this population, and it is unable to reproduce the magnitude of the mass discrepancy between the host halos of obscured and unobscured quasars. It is possible that torus and galaxy-scale obscuration (as modeled here) can contribute to this observed host mass difference to some degree, but cannot reproduce the observational results. This suggests that evolutionary models in which obscuration is an earlier stage in the lifetime of the quasar may be necessary to recover observed properties of mid-IR selected quasars.

\subsection{The Effects of Uncertainty}
Here, we address the various sources of uncertainty in our models as well as their effects on our results.
\subsubsection{Uncertainty in Scaling Relationships}
There is a degree of uncertainty inherent in the relationships that allowed us to convert our simulated $M_{\text{halo}}$ into galaxy stellar masses, and then into black hole masses \citep[e.g.,][]{most10hm-sm, haer04ej}. These uncertainties get propagated through each conversion, and they are exacerbated by the fact that these uncertainties are higher for the relationships at $z=1$ than in the local universe \citep[e.g.,][]{ haer04ej,guo10mass,most10hm-sm,behr10SMHM,lama10bh-sm}. There is also uncertainty in the observed stellar mass and black hole mass functions out at higher redshifts \citep{kell12BHMF, korm13coevolution}. To account for possible effects of uncertainty in the black hole masses at $z=1$, we probed the effect of shifting our black hole masses by $\pm 0.5$ dex for our radiation-regulated unification and galaxy-scale gas obscuration models, in accordance with the maximum error propagated through scaling relationships, as estimated in \citet{kell12BHMF}. This shift in black hole mass for our simulated quasars effectively changes the number of quasars that can be detectable after a luminosity cut is enacted, thus changing the shape of the mass distributions of our obscured and unobscured quasars. 

\begin{table*}[]
    \centering
    \caption{The effect of shifting black hole masses of our modeled quasars}
    \begin{tabular}{l c c c c}
    \hline \hline
     Model & $\log M_{\text{BH}}/M_{\odot}$ Shift & Obscured Mean  & Unobscured Mean & $f_{\text{obsc}}$ \\
      & (dex) & ($\log M_{\text{halo}}/{M_{\odot}}$) & ($\log M_{\text{halo}}/{M_{\odot}}$) &  \\
     \hline 
     Radiation-regulated & -0.5 & 12.94 & 12.92 & 0.27 \\
      & 0.0 & 12.80 & 12.75 & 0.28 \\
      & +0.5 & 12.70 & 12.59 & 0.30 \\
      \hline
     Galaxy-scale (Pannella)& -0.5 & 13.03 & 12.90 & 0.16\\
     & 0.0 & 12.91 & 12.75 & 0.14 \\
     & +0.5 & 12.80 & 12.59 & 0.12\\
     \hline
     Galaxy-scale (GRB Gauss.)& -0.5 & 12.97 & 12.83 & 0.67 \\
     & 0.0 & 12.82 & 12.65 & 0.63 \\
     & +0.5 & 12.65 & 12.57 & 0.61 \\
     \hline
     Galaxy-scale (GRB \textit{t}-dist)& -0.5 & 12.96 & 12.90 & 0.58 \\
     & 0.0 & 12.79 & 12.72 & 0.56 \\
     & +0.5 & 12.77 & 12.71 & 0.57 \\
     \hline
     Nuclear + Galaxy & -0.5 & 12.94 & 12.83 & 0.77 \\
     & 0.0 & 12.79 & 12.65 & 0.75 \\
     & +0.5 & 12.66 & 12.48 & 0.73 \\
     \hline
    \end{tabular}
    \label{tab:uncertainty}
\end{table*}

The results for this analysis are presented in Table \ref{tab:uncertainty}. Shifting the black hole masses of our sample by $\pm 0.5$ dex did not have a strong impact on the obscured fractions for any of our models. However, since shifting our black hole masses effectively changed our luminosity cut, there was a noticeable difference in the calculated obscured and unobscured mean $M_{\text{halo}}$. On average, shifting our black hole masses by -0.5 dex pushed all of the obscured and unobscured quasars to reside in more massive halos since we essentially excluded any quasars that were initially on the cusp of the luminosity cutoff. For all of the models, the shift in black hole mass of -0.5 dex produced populations of quasars whose host dark matter halos are more massive than observed, as well as obscured and unobscured populations that reside in dark matter halos of similar masses. When we shifted our black hole masses by +0.5 dex, we allowed more of our quasars to survive the luminosity cut applied. This shift had the effect of lowering the mean $M_{\text{halo}}$ for all of the obscured and unobscured populations of quasars that our models generated. Even though the mass difference between each of the obscured and unobscured populations is greater than that of the original, unshifted populations, all of the mean dark matter halos for the obscured quasars fall below that of clustering measurements.

Overall, we find that even with this systematic shift in black hole masses, our models are unable to satisfy all of the observational constraints.

\subsubsection{Uncertainty in Covering Fraction Parameterizations}
As shown above in Figure \ref{fig:ricci}, there are formal uncertainties on the $f_{\text{cov}}- \log \lambda_{\text{Edd}}$ relationship presented in \citet{ricc17edd}. In our primary analysis, we mostly considered the effect of the highly uncertain Compton-thick fraction on our models. This is because the \textit{WISE} luminosity limit eliminates the low-$\lambda_{\text{Edd}}$ quasars from our sample (as seen in Figure \ref{fig:dipompeo_ledd_dist}), so only differences in $f_{\text{cov}}$ at high-$\lambda _{\text{Edd}}$ should affect our simulated sample. However, for completeness we also explored the entire parameter space occupied by the error bounds on the original \citet{ricc17edd} relationship. We tested models that had low-$\lambda_{\text{Edd}}$ covering fractions towards the high end and the low end of the formal error bounds as well as at the same at the high-$\lambda_{\text{Edd}}$ end of the relationship. We found that none of our models that spanned the range of the \citet{ricc17edd} error bounds were able to drive significant differences between the mean halo masses of the obscured and unobscured quasar populations. The obscured fraction of quasars for most of these populations also fell below the observed obscured fraction range.

We similarly addressed the possible uncertainty in the $f_{\text{cov}}- \log M_{*}$ relationships by varying the parameterizations to cover the parameter space between the Student's $t$-distribution-derived and Gaussian PDF-derived covering fraction curves, similar to what is shown in Figure \ref{fig:ricci} for the radiation-regulated unification model. We did this to account for the fact that the shape of the underlying $N_{\text{H}}$ PDF is uncertain. We again find that there is nowhere in this parameter space that can simultaneously satify observational constraints on the dark matter halo masses for the obscured and unobscured quasars and the obscured fraction.

\subsection{Implications for Evolution}
In this work, we have explored various simple models that attempt to recover the clustering measurements of mid-IR selected quasars by characterizing quasar obscuration as a function of either $\lambda _{\text{Edd}}$ or host galaxy stellar mass. We found that these models could either satisfy dark matter halo mass measurements or the observed obscured fraction, but not both. This result strongly implies that evolution needs to be incorporated in quasar obscuration models to be able to understand the observed halo mass difference between obscured and unobscured populations of quasars. 

One commonly invoked picture of quasar evolution is that quasar activity is triggered by a dramatic event such as a merger or disk instability. The quasar then remains active in an obscured state until it rids itself of obscuring material via radiative and mechanical feedback to become unobscured \citep[e.g.,][]{sand88ULIRGs,dima05quasarfeedback,hopk08evolve,alex12bhgrowth,dipo17model,hick18obsc-review}. Qualitatively, treating the effective obscuring covering fraction as a function of time in a quasar's evolution provides a simple explanation for the fact that obscured and unobscured quasars have different observed properties such as host dark matter halo mass. \citet{dipo17model} quantitatively showed that this evolutionary sequence is able to recreate clustering measurements. The key piece to evolutionary models is understanding the timescales at which the host galaxy and the quasar/black hole evolve. The model presented in \citet{dipo17model} assumed coevolution between the host galaxy and the black hole, but the black hole grew in spurts and its growth lagged behind that of the galaxy. The implication of this is that obscured quasars host black holes that are undermassive relative to what would be expected based on their host galaxy masses. This effect coupled with a luminosity threshold is enough to drive a difference in the average host dark matter halo mass between populations of obscured and unobscured quasars. Although it has been shown that the dusty-torus does exist and that it can obscure a quasar along certain lines of sight, any torus-obscuration model needs to consider a time-dependence on the $M_{*}-M_{\text{BH}}$ relationship to be able to properly recreate observations.

Separate from host galaxy or black hole properties, \citet{powe18xrayclust} discussed the potential role of assembly bias and environment on the dark matter halo mass discrepancy between obscured and unobscured quasars. For a population of  $z \sim 0.1$, X-ray selected AGN, the model presented in \citet{dipo17model} predicted a much smaller host halo mass difference than measured. \citet{powe18xrayclust} argued that this implies that assembly bias, in which unobscured AGN reside in more recently formed halos, could  be driving a physical difference in AGN clustering. This is distinct from observed clustering differences arising as a selection effect due to the limiting luminosities of surveys. This interpretation also considers a time-dependence on obscuration, albeit on a different time scale than that in \citet{dipo17model}. Both assembly bias and event-driven evolution scenarios are viable to explain the observed clustering difference in mid-IR selected quasars on their own or in conjunction with a torus/galaxy-scale obscuration model.

\section{Summary and Conclusions}
Observational studies of quasars have shown that obscured quasars preferentially reside in higher mass dark matter halos; a result that contradicts the simplest models of unification by orientation \citep[e.g.,][]{hick11bootes_clust, dono14cluster, dipo14ir-cluster,dipo16wise-planck, dipo17qsoclust}. Recent results presented for Compton-thin AGN in \citet{ricc17edd} showed a strong relationship between the covering factor of an AGN's torus and its $\lambda _{\text{Edd}}$. Using this empirical relationship along with known $M_{\text{halo}}$ and $\lambda _{\text{Edd}}$ distributions, we constructed a simple model that sought to recreate the $M_{\text{halo}}$ difference for obscured and unobscured quasars as seen in mid-IR quasar clustering measurements. We find that our model of radiation-regulated unification is not able to recreate clustering measurements while also producing samples of quasars that have an obscured fraction that falls within observations. 

Using relationships between host galaxy gas content and stellar mass as presented in \citet{buch17mstar-nh} and \citet{pann09extinction}, it was also possible to model quasar obscuration as a function of its host galaxy's stellar mass. We find that although some of these models were able to produce host $M_{\text{halo}}$ that fell within the range of clustering measurements, they are not viable since the obscured fractions for these populations were outside the observed range \citep[e.g.,][]{trei08fobsc, poll08obscuration, dipo16torus, yan19obscWISE}. We also considered the effect of allowing our simulated quasars to be obscured by the parsec-scale dusty torus and by its host galaxy's interstellar gas. This model is able to produce a population of quasars that have an obscured fraction that falls within the observed range, but the dark matter halo mass difference between the obscured and unobscured populations is too small compared to what is calculated from mid-IR clustering measurements.

Some evolutionary paradigms of obscuration have been able to broadly recover observed dark matter halo masses of mid-IR selected quasar populations \citep[e.g.][]{dima05quasarfeedback,hopk08evolve, dipo17model, blec18agnmerge}. They assume co-evolution between the larger-scale galaxy properties and the small-scale environment of the AGN via various physical processes such as mergers or feedback. It is worth noting that even though evolutionary models have been able to reproduce dark matter halo mass measurements, they have also struggled to recover obscured fractions that fall within the range of observations \citep[e.g.,][]{dipo17model}. Here, we considered non-evolutionary physical models that describe how the properties of the galaxy or quasar could affect obscuring material on large and small scales. We implemented known empirical relationships between $M_{\text{halo}}$, galaxy mass, and SMBH mass, as well as relationships between a quasar's covering fraction and its $\lambda _{\text{Edd}}$ and between its host galaxy's stellar mass and $N_{\text{H}}$ \citep{most10hm-sm,haer04ej,pann09extinction, tink10bias,jone16sdss,ricc17edd,buch17mstar-nh,whit17obscSF}. We sought to determine if these relationships coupled with a luminosity threshold representative of the observational limitations of \textit{WISE} could recover the host $M_{\text{halo}}$ calculated via clustering measurements as well as an obscured fraction that fell within the range of observations. We found that these non-evolutionary approaches to modeling quasar evolution are not enough to be able to properly simulate observed populations of mid-IR selected quasars. We could not simultaneously recover mean $M_{\text{halo}}$ for our obscured and unobscured quasars and an obscured fraction that falls within the range of observations. The dusty torus and galaxy-scale dust and gas both likely play a role in quasar obscuration, but evolutionary models that invoke processes for AGN triggering and feedback such as event-driven radiative blowout still need to be considered to be able to model populations of observed mid-IR selected quasars.

\acknowledgments
K.E.W. acknowledges support from the Dartmouth Fellowship. R.C.H. acknowledges support from the NSF through CAREER award 1554584, and from NASA through ADAP grant number NNX16AN48G. G.T.R. acknowledges support from NASA-ADAP grant NNX17AF04G. A.D.M. acknowledges support from NSF grant AST-1616168. Thank you to the anonymous referee whose input helped make this a stronger paper. We also thank Laurrane Lanz and Alberto Masini for their constructive input.

\bibliography{whalen_research}
\end{document}